\documentclass[prd,amsmath,amssymb,superscriptaddress,floatfix,nofootinbib,10pt]{revtex4}
\usepackage{times}
\usepackage{amssymb,amsbsy,amsmath,amsfonts}
\usepackage{graphicx}
\usepackage{float}
\usepackage{color}
\usepackage{morefloats}
\usepackage{rotating}
\usepackage{srcltx}
\usepackage{slashed}
\usepackage{multirow}
\usepackage{verbatim}
\usepackage{hyperref}
\usepackage{tabularx}
\usepackage{bm}
\allowdisplaybreaks[4]
\usepackage{bbding}
\usepackage{threeparttable}
\newcommand{\PreserveBackslash}[1]{\let\temp=\\#1\let\\=\temp}
\newcolumntype{C}[1]{>{\PreserveBackslash\centering}p{#1}}
\newcolumntype{R}[1]{>{\PreserveBackslash\raggedleft}p{#1}}
\newcolumntype{L}[1]{>{\PreserveBackslash\raggedright}p{#1}}

\begin{document}

\title{Evolution of compact states to molecular ones with coupled channels: The case of the $X(3872)$ }

\begin{abstract}
  We study the molecular probability of the $X(3872)$ in the $D^0 \bar D^{*0}$ and $D^+ D^{*-}$ channels in several scenarios. One of them   assumes that the state is purely due to a genuine nonmolecular component. However,   it gets unavoidably dressed by the meson components to the point that in the limit of zero binding of the $D^0 \bar D^{*0}$ component   becomes purely molecular. Yet, the small but finite binding   allows for a nonmolecular state when the bare mass of the genuine state approaches the $D^0 \bar D^{*0}$ threshold, but, in this case the system develops a small scattering length and a huge effective range for this channel in flagrant disagreement with present values of these magnitudes.   Next we discuss the possibility to have hybrid states stemming from the combined effect of a genuine state and a reasonable direct interaction between the meson components, where we find cases in which the scattering length and effective range are still compatible with data, but even then the molecular probability is as big as $95 \%$. Finally, we perform the calculations when the binding stems purely from the direct interaction between the meson-meson components.  In summary we conclude, that while present data definitely rule out the possibility of a dominant nonmolecular component, the precise value of the molecular probability requires   a more precise determination of the scattering length and effective range of the $D^0 \bar D^{*0}$ channel, as well as the measurement of these magnitudes for the  $D^+ D^{*-}$ channel which have not been determined experimentally so far.
\end{abstract}

%\pacs{13.75.Ev,12.39.Fe,21.30.Fe}
%\keywords{}

\date{\today}
\author{Jing Song}
\email[E-mail me at: ]{Song-Jing@buaa.edu.cn}
\affiliation{School of Physics, Beihang University, Beijing, 102206, China}
\affiliation{Departamento de Física Teórica and IFIC, Centro Mixto Universidad de Valencia-CSIC Institutos de Investigación de Paterna, 46071 Valencia, Spain}

\author{L.R.Dai}
\email[E-mail me at: ]{ dailianrong@zjhu.edu.cn}
\affiliation{School of science, Huzhou University, Huzhou, 313000, Zhejiang, China}
\affiliation{Departamento de Física Teórica and IFIC, Centro Mixto Universidad de Valencia-CSIC Institutos de Investigación de Paterna,  46071 Valencia, Spain}

\author{ E.Oset}
\email[E-mail me at: ]{oset@ific.uv.es}
\affiliation{Departamento de Física Teórica and IFIC, Centro Mixto Universidad de Valencia-CSIC Institutos de Investigación de Paterna, 46071 Valencia, Spain}

\maketitle

\section{introduction}
The discovery of hadronic states of exotic nature, challenging the standard $q\Bar{q}$ nature for mesons and $qqq$ nature for baryons has lead to a revival of hadron physics and many review papers have been devoted to study these new systems~\cite{Esposito:2014rxa,Lebed:2016hpi,Chen:2016qju,Guo:2017jvc,Kalashnikova:2018vkv,Yamaguchi:2019vea,Brambilla:2019esw,Guo:2019twa,Chen:2022asf,Mai:2022eur}. One of the recurring questions about such systems is whether they are better explained in tems of compact tetraquarks or pentaquarks, or they follow a different patern as molecular states of meson-meson for mesonic states or meson-baryon for the baryonic states. \\

Referring to the last three years (earlier references can be found in the review papers cited), there is a large amount of papers discussing the nature of the states, some claiming a molecular nature of the $D^0 \bar D^{*0}$ and  $D^+ D^{*-}$ (and $cc$) type ~\cite{Braaten:2020nmc,Liu:2020tqy,Meng:2020cbk,Dong:2021juy,Wu:2021udi,Gordillo:2021bra,Dong:2021bvy,Meng:2021kmi,Kamiya:2022thy,Lin:2022wmj,Ji:2022uie,Wang:2022qxe,Wang:2022xga,Kinugawa:2023fbf,Yang:2023mov,Wu:2023rrp,Peng:2023lfw,Terashima:2023tun}, and others claiming a compact tetraquark state \cite{Shi:2021jyr,Esposito:2021vhu,Huang:2021poj,Chen:2022ddj,Sharma:2022ena}. Some people advocate a mixture of the two structures \cite{Lebed:2022vks,Wang:2023ovj} and discussions around this possible scenario are done in~\cite{Kang:2016jxw,Baru:2010ww}. Also much work has  been devoted to show the relevance of studying the $X(3872)$ in $pp$ and heavy ion collisions as a means to further learn about the structure of the state \cite{Zhang:2020dwn,Wu:2020zbx,Braaten:2020iqw,Xu:2021drf,Esposito:2021vhu}. The possibility of learning about this structure by looking at the $X(3872)$ in a nuclear medium has also been discussed in \cite{Albaladejo:2021cxj}.\\

As one can see, the majority of papers advocate a molecular structure, but other works find support for the compact tetraquark nature. The fact that the state is so close to the $D^0\Bar{D}^{*0}$ threshold favors the molecular structure, and this and other reasons have been used to support the molecular structure. However, as we shall see, the proximity of the state to a threshold does not guarantee by itself that the state is of molecular nature, although certainly is favors it. A discussion on this issue for the $T_{cc}(3875)$ is done in~\cite{Dai:2023kwv}.\\

The purpose of the present work is to shed light on the issue of the $X(3872)$ compositeness. For this purpose we start from a genuine nonmolecular state, which couples to  $D^0\Bar{D}^{*0}$, the channel where it is observed, and then assume that by itself it provides a  bound state in the  $D^0\Bar{D}^{*0}$ component. What we observe is that unavoidably the state develops a molecular component and we evaluate its probability, which becomes unity in the limit of zero binding. The question arises about the `scale' of what small means in the real case and we  investigate  that in terms of the bare mass of the genuine state. We conclude that it is perfectly possible to have a bound state produced which is still of nonmolecular nature if the genuine state mass is sufficiently close to the threshold. Yet, one pays a price for this, since then, the scattering length of  $D^0\Bar{D}^{0}$ becomes small and the effective range grows indefinitely. With present values of this information one can then rule that scenario, concluding the unavoidable molecular nature of the state.\\

\section{formalism}
Let us start with a state of nonmolecular nature with a bare mass $m_R$, which couples to  $\Bar{D}^0{D}^{*0}$,  $D^-{D}^{*+}$. The results of this work come from the mass of these states and we can ignore the complex conjugate component. However, following the isospin assignment of the PDG~\cite{ParticleDataGroup:2022pth}  we will assume that the state has $I=0$ (some isospin breaking will appear as a consequence of the different mass of $\Bar{D}^0\Bar{D}^{*0}$,  $D^-{D}^{*+}$). With the isospin multiplets ($D^+,~-{D}^{0}$), ( $\Bar{D}^0,~{D}^{-}$), ( ${D}^{*+},~-{D}^{*0}$), ( $\Bar{D}^{*0},~{D}^{*-}$) the isospin zero state is given by,

\begin{align}\label{3_1}
|{D}^*\Bar{D},~I=0\rangle=\frac{1}{\sqrt{2}}(D^{*0}\Bar{D}^0+D^{*+}{D}^-)
\end{align}

\begin{figure}[H]
    \centering
\includegraphics[width=0.5\textwidth]{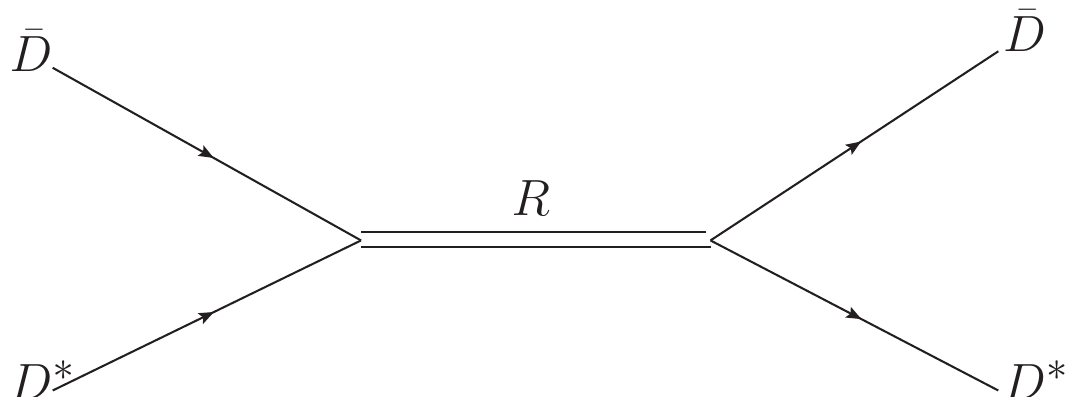}
    \caption{${D}^*\Bar{D}$ amplitude from the genuine resonance $R$.}
    \label{FIG1}
\end{figure}

\begin{figure}[H]
    \centering
\includegraphics[width=0.25\textwidth]{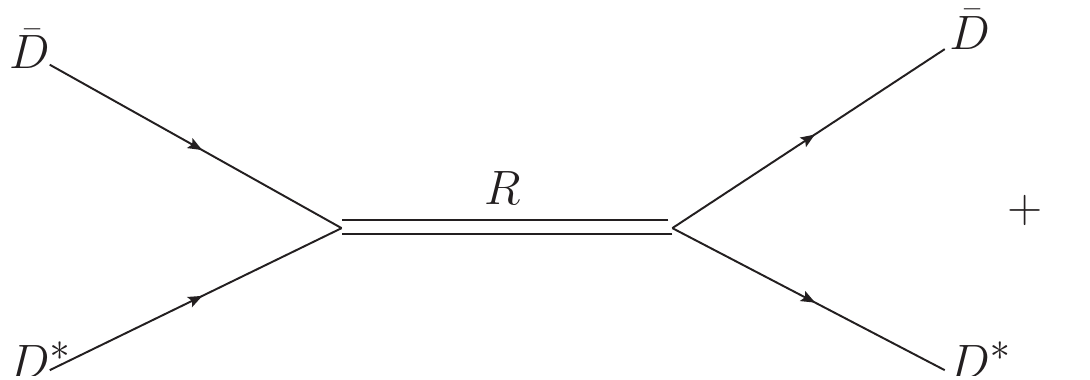} 
\includegraphics[width=0.28\textwidth]{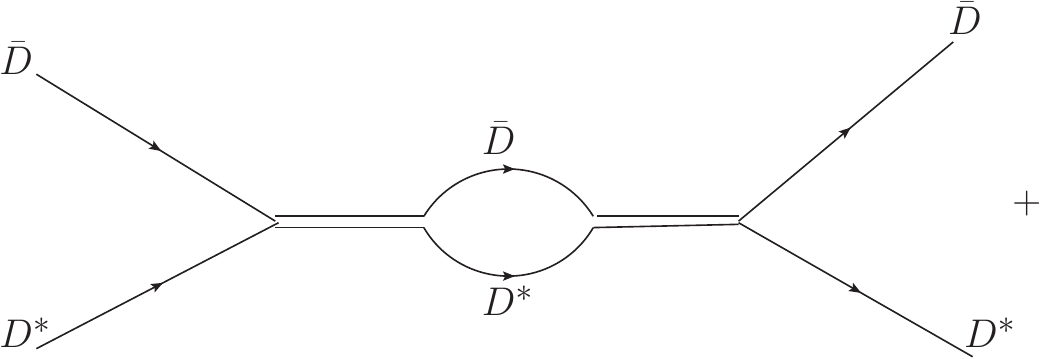}
\includegraphics[width=0.38\textwidth]{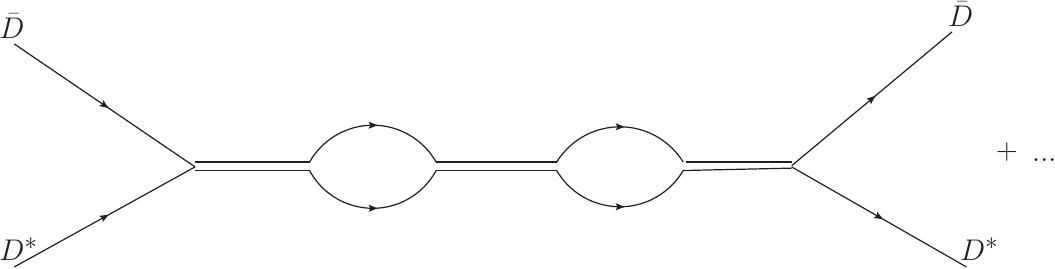}
    \caption{Dressing of amplitude of Eq.~(\ref{3_2}) by the ${D}^*\Bar{D}$ selfenergy.}
    \label{FIG2}
\end{figure}

Let us represent the genuine state coupling to this $I=0$ state by  
\begin{align}\label{3_2}
t_{{D}^*\Bar{D}}(I=0)= \frac{\Tilde{g}^2}{s-s_R}
\end{align}
represented by Fig.~{\ref{FIG1}}, 
where $s_R$ represents the mass of this state previous to the unavoidable dressing by the meson-meson component, and $\Tilde{g}^2$ provides the strength of this interaction.
From now on we work with the coupled channels  ${D}^{*0}\Bar{D}^{0}$ (1), ${D}^{*+}{D}^{-}$(2)   and the amplitude of Eq.~(\ref{3_2}) in coupled channels becomes

\begin{align}\label{3_3}
\centering
\Tilde{V}_R=
\left(
  \begin{array}{cc}
    \frac{1}{2}  & \frac{1}{2}\\
    \frac{1}{2}  & \frac{1}{2}\\
  \end{array}
\right)~\frac{\Tilde{g}^2}{s-s_R} \equiv
\left(
  \begin{array}{cc}
    \frac{1}{2} V_R & \frac{1}{2} V_R\\
    \frac{1}{2} V_R  & \frac{1}{2} V_R\\
  \end{array}
\right),\qquad \text{with} ~~V_R=\frac{\Tilde{g}^2}{s-s_R}.
\end{align}
The amplitude of Eq.~(\ref{3_3}) is not unitary. Unitary is accomplished by dressing the amplitude of Fig.~\ref{FIG1} with the selfenergy of the $\Bar{D}D^+$ components, as shown in Fig.~\ref{FIG2}. 
Let $T$ be the unitary amplitude which is given by, 
\begin{align}\label{3_4}  T=\Tilde{V}_R+\Tilde{V}_RG\Tilde{V}_R+\Tilde{V}_RG\Tilde{V}_RG\Tilde{V}_R+...=\Tilde{V}_R+\Tilde{V}_RGT
\end{align}
with $G$ the diagonal loop matrix
\begin{align}
\centering
G=
\left(
  \begin{array}{cc}
     G_{{D}^{*0}\bar{D}^{0}} & 0 \\
    0  &  G_{{D}^{*+}{D}^{-}}  \\
  \end{array}
\right).\nonumber
\end{align}
The function $G$ is regularized with a cut off $q_\mathrm{max}$, and we find 
\begin{align}\label{3_5}
   G_i(s)=\int_{|\textbf{q}|< q_{\mathrm{max}}}&\frac{d^3\textbf{q}}{(2\pi)^3}\frac{w_1^{(i)}+w_2^{(i)}}{2w_1^{(i)}w_2^{(i)}}
   \times \frac{1}{s-(w_1^{(i)}+w_2^{(i)})^2+i\epsilon}
\end{align}
where $i={D}^{*0}\bar{D}^{0},~{D}^{*+}{D}^{-}$ and $w_1,~~w_2$ are given by $w_1=\sqrt{\textbf{q}^2+m^2_{{D}^{*}}}~~w_2=\sqrt{\textbf{q}^2+m^2_{\Bar{D}}}$.
Eq.~(\ref{3_4}) gives 
\begin{align}\label{3_6} 
T=[1-\Tilde{V}_RG]^{-1}\Tilde{V}_R
\end{align}
and provides the scattering matrix, $T$, for the two channels considered. We easily find analytically that
\begin{align}\label{3_7}
\centering
T=
\frac{1}{\mathrm{DET}}
\left(
  \begin{array}{cc}
    \frac{1}{2} V_R & \frac{1}{2} V_R\\
    \frac{1}{2} V_R  & \frac{1}{2} V_R\\
  \end{array}
\right),
\end{align}
with DET being the determinant of $(1-\Tilde{V}_RG)$

\begin{align}\label{3_8}
 \mathrm{DET}=  1- \frac{1}{2}V_RG_1-\frac{1}{2}V_RG_2
\end{align}
If we decide to have a bound state at $s_0$, the $T$ will have a pole at that energy implying
\begin{align}\label{3_9}
 1- {\frac{1}{2}V_RG_1(s_0)-\frac{1}{2}V_RG_2(s_0)}=0
\end{align}
from where, given $s_R$ we can obtain $\Tilde{g}^2$ as 
\begin{align}\label{3_10}
\tilde{g}^2 = \frac{{s}-{s_R}}{\frac{1}{2}G_1+\frac{1}{2}G_2}\Bigg|_{s_0}.
\end{align}

\subsection{Couplings and probabilities}
The couplings to the state for the two channels are given by 
\begin{align}\label{3_11}
g_1^2 =\lim(s-s_0)T_{11};&\qquad g_2^2 =\lim(s-s_0)T_{22}\\ \nonumber
g_2&=g_1\lim(s-s_0)\frac{T_{21}}{T_{11}}
\end{align}
From Eqs.~(\ref{3_7}),~(\ref{3_8}) using L'Hospital's rule we easily find 
\begin{align}\label{3_12}
g_1^2 =\frac{\frac{1}{2}{\tilde{g}^2}}{1-\frac{1}{2}{\tilde{g}^2}\frac{\partial}{\partial s}(G_1+G_2)}\Bigg|_{s_0}; \qquad g_2 =g_1 
\end{align}
The fact that $g_2 =g_1 $ does not exactly imply that we have $I=0$, according to Eq.(~\ref{3_1}). Indeed, in strong interactions where the isospin manifest itself, given the short range of the interaction, what matters is the wave function at the origin and this is given by~\cite{Gamermann:2009uq}
\begin{align}\label{3_13}
    \Psi_1(r=0)=g_1G_1(s_0);\qquad \Psi_2(r=0)=g_2G_2(s_0)
\end{align}
and, since $G_1,~G_2$ are different due to the different mass of the channels, $\Psi_1$ and $\Psi_2$ are a bit different. The discussion about isospin violation for the $X(3872)$ was already done in Ref.~\cite{ Gamermann:2009fv}. 

Once we have the couplings $g_1^2$, $g_2^2$, we can calculate the probabilities to have the ${D}^{*0}\bar{D}^{0}$ and ${D}^{*+} {D}^{-}$ in the wave function of the $X(3872)$ as~\cite{Gamermann:2009uq,Hyodo:2013nka}.
\begin{align}\label{3_14}
P_1  &= -g_1^2 \frac{\partial G_1}{\partial s}|_{s_0}
= -\frac{\frac{1}{2}{\tilde{g}^2}\frac{\partial G_1}{\partial s}}{1-\frac{1}{2}{\tilde{g}^2}\frac{\partial}{\partial s}(G_1+G_2)}\Bigg|_{s_0}\\ \nonumber
P_2  &= -g_2^2 \frac{\partial G_2}{\partial s}|_{s_0}
= -\frac{\frac{1}{2}{\tilde{g}^2}\frac{\partial G_2}{\partial s}}{1-\frac{1}{2}{\tilde{g}^2}\frac{\partial}{\partial s}(G_1+G_2)}\Bigg|_{s_0}
\end{align}
Since the $X(3872)$ is closer to the channel $1$ threshold (${D}^{*0}\bar{D}^{0}$) and $\frac{\partial G_1}{\partial s} \to \infty$ when $s_0\to s_\mathrm{th1}$ we immediately find that\\

(1)
When $\tilde{g}^2 \to 0$,\quad  $P_1\to 0$, $P_2\to 0$, genuine state.\\

(2)
When $\tilde{g}^2 \to \infty$,\quad  $P_1+P_2= 1$, completely molecular.\\

(3)
When  $s_0\to s_\mathrm{th1}$, $\frac{\partial G_1}{\partial s} \to \infty$, and $\frac{\partial G_2}{\partial s} \to \mathrm{finite}$, \quad  $P_1\to 1$, $P_2\to 0$, completely molecular state dominated by the ${D}^{*0}\bar{D}^{0}$ component.\\

Let us stress again that even if  $P_1\to 1$, $P_2\to 0$, in strong interaction of zero range what matters is the wave function at the origin and the ${D}^{*0}\bar{D}^{0}$ and ${D}^{*+}{D}^{-}$ components become equally important~\cite{Gamermann:2009fv}.

\subsection{Inclusion of direct interaction between channels}
As shown in~\cite{Gamermann:2009uq} and the different pictures claiming a molecular nature for the $X(3872)$, there is a direct interaction between the ${D}^{*}\bar{D}$ components due to meson exchange. There are differences between the different models but all them conclude that the interaction is attractive. In the local hidden gauge approach the interaction comes from the exchange of vector mesons~\cite{Bando:1987br,Harada:2003jx,Meissner:1987ge,Nagahiro:2008cv}. The vertices needed for this interaction can be obtained from the Lagrangian
\begin{align}\label{3_15}
    \mathcal{L} = -ig' \langle [P,\partial_\mu P] V^\mu\rangle,
\end{align}
with $g'=\frac{m_\mathrm{v}}{2f_\pi}$, $m_\mathrm{v}=800$~MeV, $f_\pi=93$~MeV,
where $P$, $V$ are the $q\Bar{q}$ matrices expressed in terms of pseudoscalars, vector mesons~\cite{Ikeno:2020mra} and $\langle ~~ \rangle$ is the trace of the matrices.

With the exchange of $\rho,~w$ mesons we obtain an interaction
\begin{align}\label{3_16}
\centering
\Tilde{V}=
\left(
  \begin{array}{cc}
    \frac{1}{2} V & \frac{1}{2} V\\
    \frac{1}{2} V & \frac{1}{2} V\\
  \end{array}
\right)
\end{align}
with  
\begin{align}\label{3_17}
\frac{1}{2}V =-g'^2\frac{4 ~m_{D^{*0}}m_{D^0}}{m^2_V}.
\end{align}
calculated at threshold. We use this interaction as a scale of the interaction, keeping in mind that the binding is tied to the interaction but also to $q_\mathrm{max}$ in the $G$  function of Eq.~(\ref{3_5}) (in one channel  one has $T=(V^{-1}-G)^{-1}$ and the pole appear at $V^{-1}-G=0$, hence, changes of $V^{-1}$ can be accommodated by changes in $q_\mathrm{max}$ and  vice versa. We shall play with this  flexibility by showing results with two different values of $q_\mathrm{max}$). The formulas obtained before for $\Tilde{g}^2$, the couplings and probabilities are  trivially modified by changing 
\begin{align}\label{3_18}
V_R \to V_R+\beta V
\end{align}
where $\beta$, for the sake of showing which is the result of adding a direct interaction, will be taken for each $q_\mathrm{max}$ such that we barely do not bind the state with only the $\beta V$ interaction, in other words we would get the binding energy zero with this value of $\beta$ for the chosen value of $q_\mathrm{max}$.

\subsection{Scattering length and effective range}
With the normalization of $G$ in Eq.~(\ref{3_5}) and our choice of $V_R,~V$, we have the relationship between our $T$ matrix and the scattering matrix used in Quantum mechanics as 
\begin{align}\label{3_19}
T= (-8\pi\sqrt{s} ) f^{QM} \simeq
 (-8\pi\sqrt{s} ) \frac{1}{-\frac{1}{a}+\frac{1}{2}r_0 k^2-ik}
\end{align}
with $a$ and $r_0$ the scattering length and effective range. We have these magnitudes defined for every channel from $T_{11}$ and $T_{22}$. Then we easily find 
\begin{align}\label{3_20}
T_{jj}&= \frac{1}{\frac{s-{s_R}}{\frac{1}{2}[{\tilde{g}^2}+\beta V({s-{s_R}})]}-G_1-G_2},\qquad j=1,2     
\end{align}
from where 
\begin{align}\label{3_21}
{-\frac{1}{a}+\frac{1}{2}r_0 k^2-ik} = (-8\pi\sqrt{s} ) (T_{jj})^{-1} = (-8\pi\sqrt{s} ) \Bigg\{\frac{s-{s_R}}{\frac{1}{2}[{\tilde{g}^2}+\beta V({s-{s_R}})]}-G_1-G_2\Bigg\}. 
\end{align}
We immediately see that $-ik$ for the two channels comes automatically from $i8\pi\sqrt{s}~ \mathrm{Im} G_j$ since $\mathrm{Im} G=\frac{-1}{8\pi\sqrt{s}}k$.
 Then we  obtain 
\begin{align}\label{3_22}
-\frac{1}{a_1}  &=  (-8\pi\sqrt{s} ) \Bigg[{\frac{s-{s_R}}{\frac{1}{2}[{\tilde{g}^2}+\beta V({s-{s_R}})]}-\mathrm{Re}G_1-G_2}\Bigg]\Bigg|_\mathrm{s_{th1}},
\end{align}

\begin{align}\label{3_23}
r_{0,1} &= 2\frac{\sqrt{s}}{\mu_1}\frac{\partial}{\partial{s}} \Bigg\{(-8\pi\sqrt{s} ) \Bigg[\frac{s-{s_R}}{\frac{1}{2}[{\tilde{g}^2}+\beta V({s-{s_R}})]}-\mathrm{Re}G_1-G_2\Bigg]\Bigg\}\Bigg|_\mathrm{s_{th1}},
\end{align}

\begin{align}\label{3_24}
-\frac{1}{a_2}  &=  (-8\pi\sqrt{s} ) \Bigg[{\frac{s-{s_R}}{\frac{1}{2}[{\tilde{g}^2}+\beta V({s-{s_R}})]}-\mathrm{Re}G_2-G_1}\Bigg]\Bigg|_\mathrm{s_{th2}},
\end{align}

\begin{align}\label{3_25}
r_{0,2} &= 2\frac{\sqrt{s}}{\mu_2}\frac{\partial}{\partial{s}} \Bigg\{(-8\pi\sqrt{s} ) \Bigg[\frac{s-{s_R}}{\frac{1}{2}[{\tilde{g}^2}+\beta V({s-{s_R}})]}-\mathrm{Re}G_2-G_1\Bigg]\Bigg\}\Bigg|_\mathrm{s_{th2}},
\end{align}
with $\mu_i$ the reduced mass of the channel.\\

Note that in Eqs.~(\ref{3_22})~(\ref{3_23}) $G_2$ is real. However, Eqs.~(\ref{3_24})~(\ref{3_25}) $G_1$ is complex since it is evaluated at the second threshold $\mathrm{s_{th2}}>\mathrm{s_{th1}}$. Then $a_2,~r_{0,2}$ will be complex, while $a_1,~r_{0,1}$ will be real in the approximation that we do of neglecting the $D^*$ width. The changes introducing the $D^*$ width are small as seen in the study of the $T_{cc}(3875)$~\cite{Dai:2023cyo}, definitely much smaller than the differences that we find for different scenarios of the structure of the $X(3872)$.

The formulas in Eqs.~(\ref{3_22})-(\ref{3_25}) can equally be used for the case neglecting the direct interaction between the components, simply setting $\beta=0$.

\section{results}

We take the masses from the PDG as 
\begin{align}\label{3_26}
&m_{D^{*0}} = 2006.85~\mathrm{MeV}, \quad
m_{{D}^0} = 1864.84~\mathrm{MeV}, \quad
m_{D^{*+}} = 2010.26~\mathrm{MeV}, \quad
m_{{D}^+} = 1869.66~\mathrm{MeV}, \\ \nonumber
&m_{X(3872)} = 3871.65~\mathrm{MeV}, \quad
m_{D^{*0}} + m_{{D}^0} = 3871.69~\mathrm{MeV}, \quad
m_{D^{*+}} + m_{{D}^-}  = 3879.92~\mathrm{MeV}.
\end{align}
as we can see, the $X(3872)$ state is barely $40$~KeV below the $D^{*0}\Bar{D}^0$ threshold, extremely weakly bound.
We take different values of $ \sqrt{s}_R= \sqrt{s}_\mathrm{th1}+\Delta \sqrt{s}_R$ and plot the results for different values of $\Delta \sqrt{s}_R$.

\begin{figure}[H]
    \centering
\includegraphics[width=0.4\textwidth]{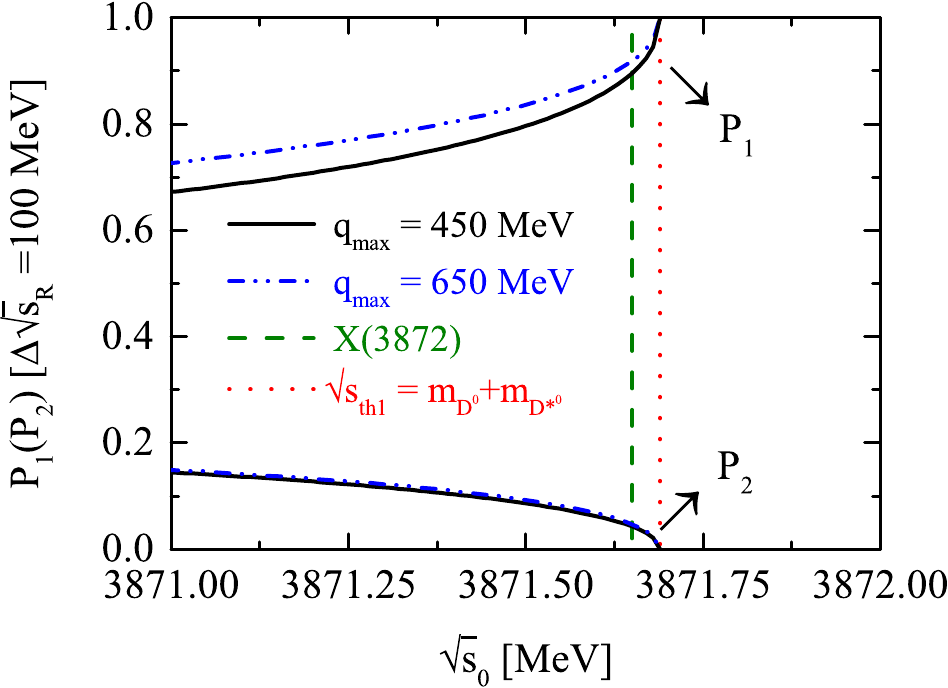}
\includegraphics[width=0.4\textwidth]{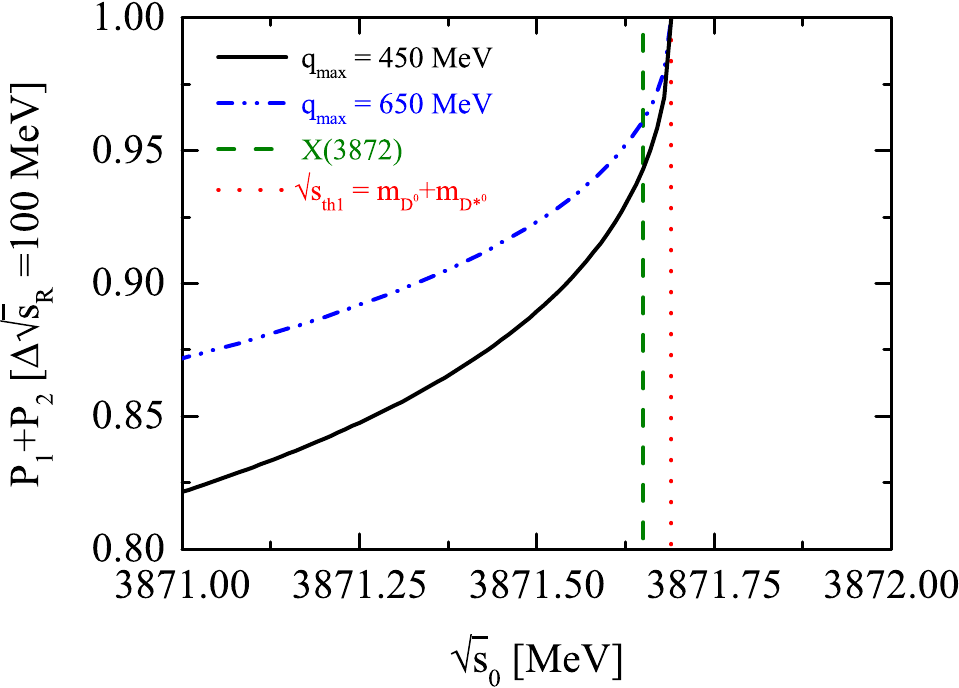} 
\caption{The evolution of $P$ with $\sqrt{s}$. The red lines are the threshold of $D^0D^{*0}$, the olive-dashed lines are the bound state of $X(3872)$, and the black and blue curves present the behaviors of $q_\mathrm{max}=450~$MeV and $q_\mathrm{max}=650~$MeV, respectively.}
    \label{FIG3}
\end{figure}

\begin{figure}[H]
    \centering
\includegraphics[width=0.4\textwidth]{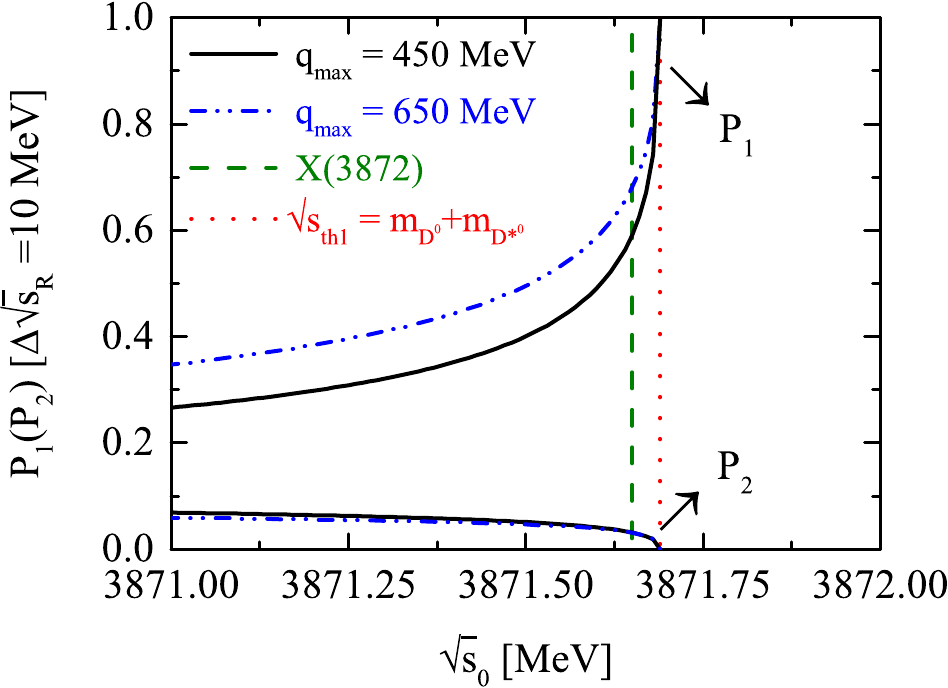}
\includegraphics[width=0.4\textwidth]{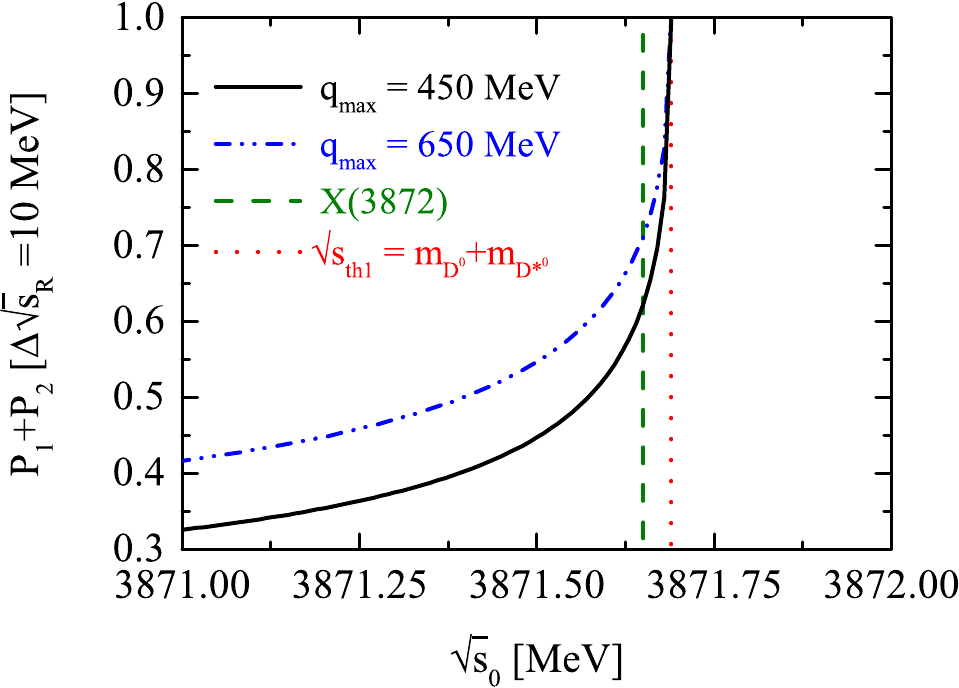}    
\caption{The labels are same as Fig.~\ref{FIG3}.}
    \label{FIG5}
\end{figure}

\begin{figure}[H]
    \centering
\includegraphics[width=0.4\textwidth]{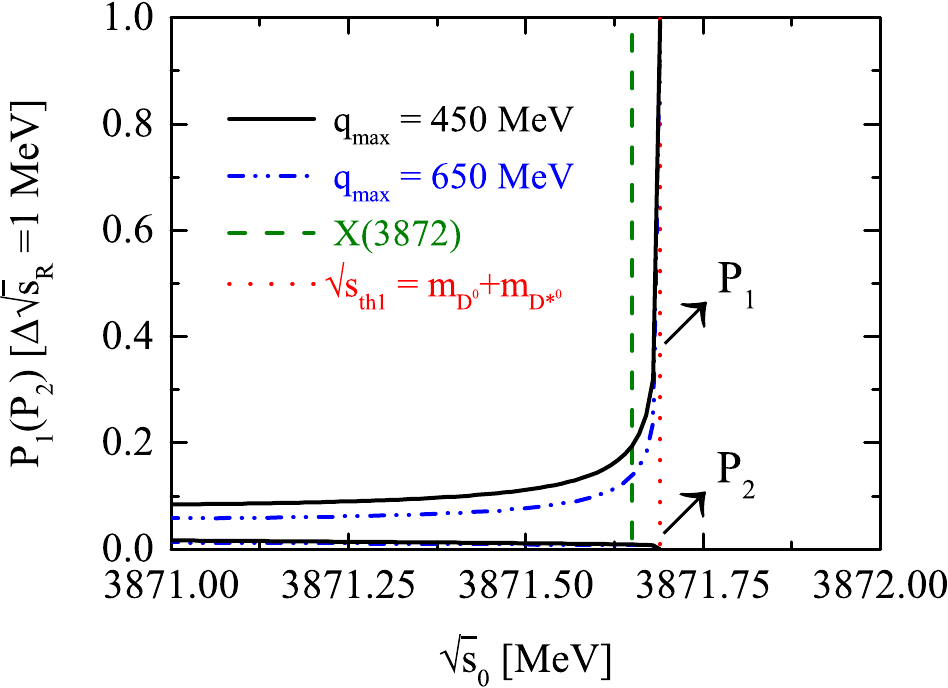}
\includegraphics[width=0.4\textwidth]{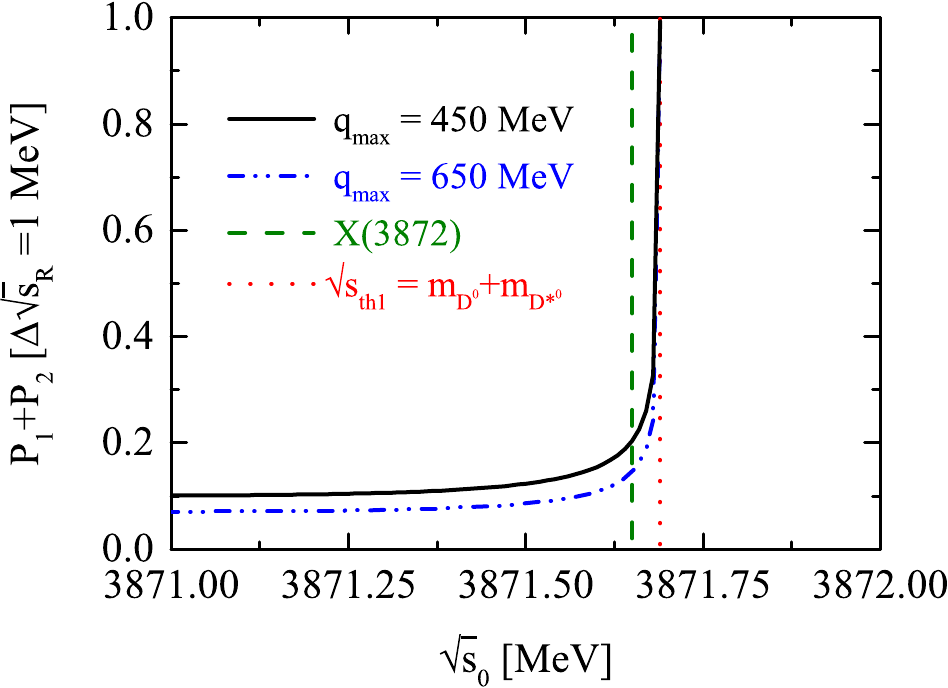}    
\caption{The labels are same as Fig.~\ref{FIG3}.}
    \label{FIG7}
\end{figure}

\begin{figure}[H]
    \centering
\includegraphics[width=0.4\textwidth]{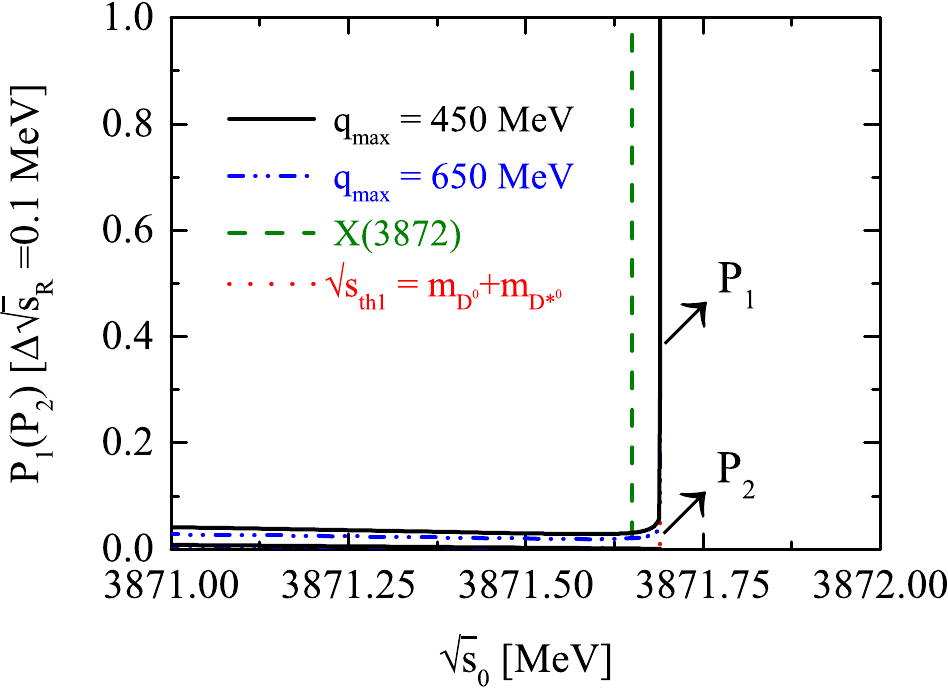}
\includegraphics[width=0.4\textwidth]{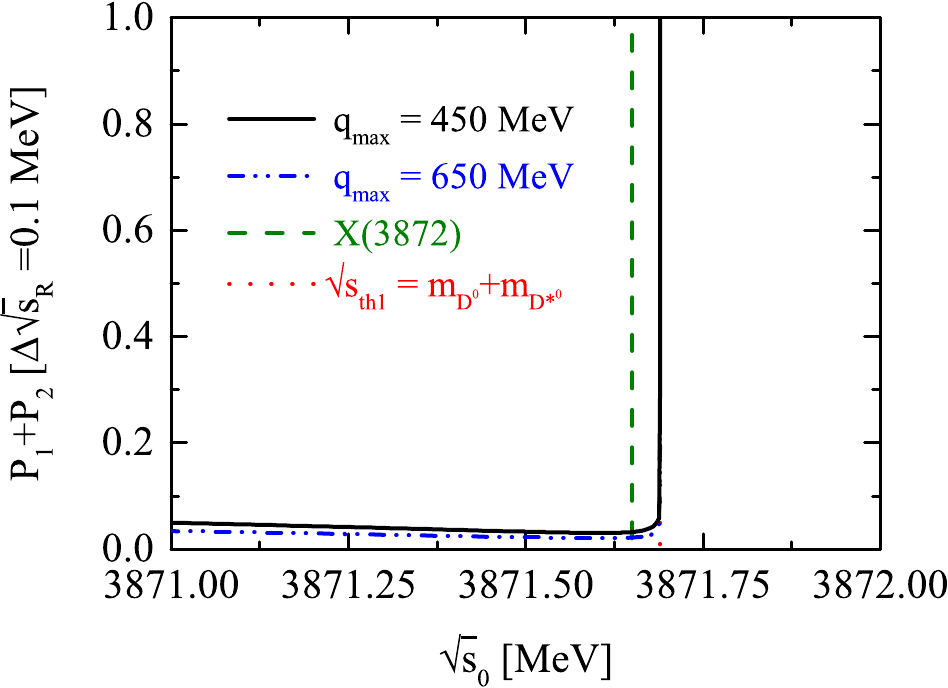}    
\caption{The labels are same as Fig.~\ref{FIG3}.}
    \label{FIG9}
\end{figure}

In Figs.~\ref{FIG3} to \ref{FIG9} we neglect the direct interaction of the mesons, $(\beta=0)$ in Eq.~(\ref{3_18}). 
In Fig.~\ref{FIG3} left we see the results for $P_1,~P_2$ for $\Delta \sqrt{s}_R=100$~MeV, meaning that the mass of the genuine state is $100$~MeV about the ${D^{*0}}\Bar{D}^0$ threshold. What we observe is that as $\sqrt{s_0}$ goes to the ${D^{*0}}\Bar{D}^0$ threshold, $P_1$ goes to 1and $P_2$ goes to zero, as we expected. We also see that at the energy of $X(3872)$ the probability $P_1\sim 0.9$ and $P_2\sim 0.05$ depending a bit on the choice of  $q_\mathrm{max}$. In Fig.~\ref{FIG3} right we  show $P_1+P_2$ to see the convergence to 1 of the sum of the two components when we approach $\mathrm{s_{th1}}$. The total molecular probability is around $0.95$ at the $X(3872)$ energy. This means that we started from a state that was purely nonmolecular but it got dressed by the meson-meson component to the point that this component assumes most of the probability in the wave function of the state.

In Fig.~\ref{FIG5} left we repeat the procedure with a $m_R$ mass closer to the ${D^{*0}}\Bar{D}^0$ threshold, only $10$~MeV above. We observe that the theorem of $P_1\to 1$ at threshold holds again, but at the pole of the $X(3872)$ $P_1\sim 0.6-0.7$ and $P_2\approx0.03$. In Fig.~\ref{FIG5} right we see that $P_1+P_2\sim 0.6-0.7$. In this case the molecular probability is smaller than before, indicating that if the bare mass of the genuine state is closer to threshold the amount of induced molecular component is smaller, but still sizeable.

In Fig.~\ref{FIG7} left we repeat the procedure with $\Delta \sqrt{s}_R=1$~MeV. In this case at $\sqrt{s}_0$ the value of $P_1\sim0.15-0.2$ and the one of $P_2\approx0.01$. The sum $P_1+P_2$ is shown in Fig.~\ref{FIG7} right and we see that $P_1+P_2\sim0.15-0.2$, very small.

Finally in Fig.~\ref{FIG9} we show the results for $\Delta \sqrt{s}_R=0.1$~MeV. In this case we see that the $P_1+P_2$ is around $0.02$, indicating that the induced molecular component is negligible.

The conclusion from all these results is that the binding energy by itself does not give us the molecular probability and it is possible to have a very small binding and still have a negligible molecular component. 

However, we can see what happens with $a$ and $r_0$ in those cases. This is shown in Table~\ref{value_a1_sR}. What we can see is that the scattering lengths $a_1,~a_2$ become small, and most important, the value of the effective range become extremely large, bigger than $600$~fm in size for the case of $\Delta\sqrt{s}_R=0.1$~MeV when we had a negligible molecular component. This should be contrasted with present experimental values. From the study of the line shape in $D\Bar{D}\pi$ production of LHCb~\cite{LHCb:2020xds} the authors of Ref.~\cite{Esposito:2021vhu} induce 
\begin{align}\label{3_27}
    r_{0,1} =-5.34~\mathrm{fm}.
\end{align} 
However, the authors of~{\cite{Baru:2021ldu}} redo an analysis of the data and after subtracting the  contribution from the second channel they get a value of around $-3.78$~fm. Different corrections from unknown elements in the theoretical framework reduce the radius $r_{0,1}$ to~{\cite{Baru:2023commu}}
\begin{align}\label{3_28}
    -2.78~\mathrm{fm}~<~ r_{0,1}~<1~\mathrm{fm}.
\end{align} 
The value extracted for $a_1$, after accounting for the different prescription in~{\cite{Baru:2021ldu}} ($\frac{1}{a}$ instead of $-\frac{1}{a}$ in our case), is
\begin{align}\label{3_29}
a_1\approx28~\mathrm{fm}.
\end{align} 
which is very small and has large uncertainties due to uncertainties in the binding.

The discrepancy of the results in Table~\ref{value_a1_sR}, when one has a small molecular component, with the experimental data on  $a_1$ and $r_{0,1}$ is large, enough to discard this scenario. The values obtained for these magnitudes for $\sqrt{s}_R=100$~MeV, would be basically acceptable, but in this case we found that the molecular component is close to unity.

\subsection{Results with a mixture of genuine state and direct meson-meson interaction}
\begin{figure}[H]
    \centering
\includegraphics[width=0.4\textwidth]{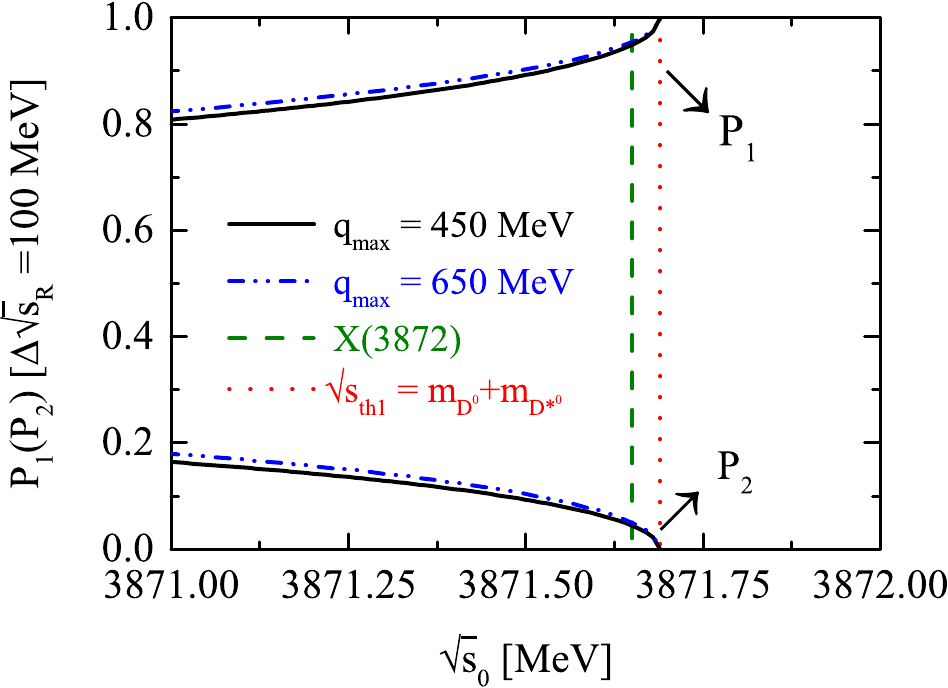}
\includegraphics[width=0.4\textwidth]{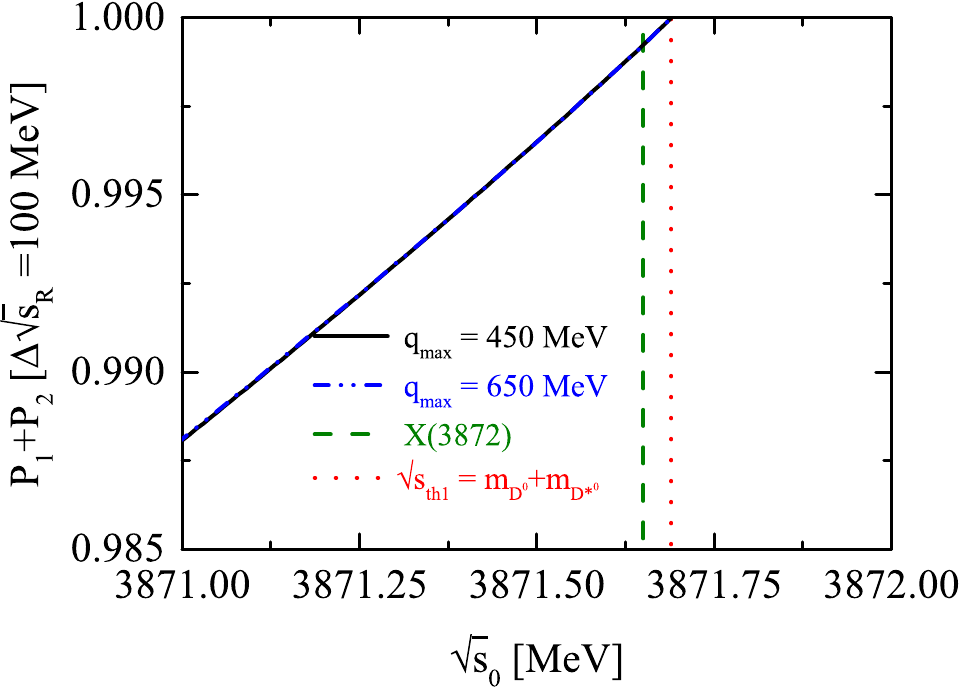}    
\caption{The labels are same as Fig.~\ref{FIG3}.}
    \label{FIG11}
\end{figure}

\begin{figure}[H]
    \centering
\includegraphics[width=0.4\textwidth]{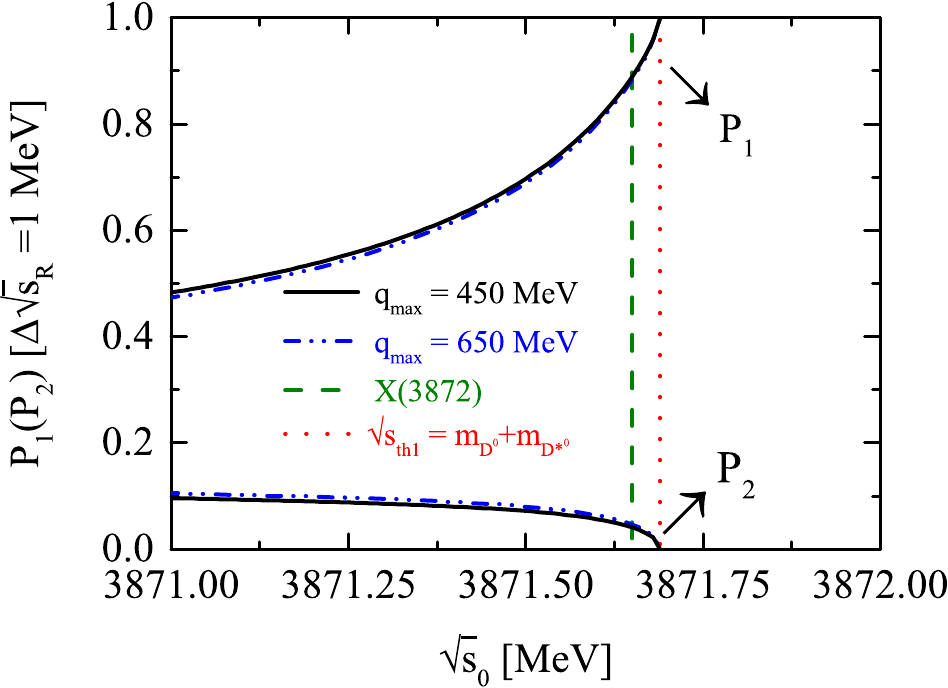}
\includegraphics[width=0.4\textwidth]{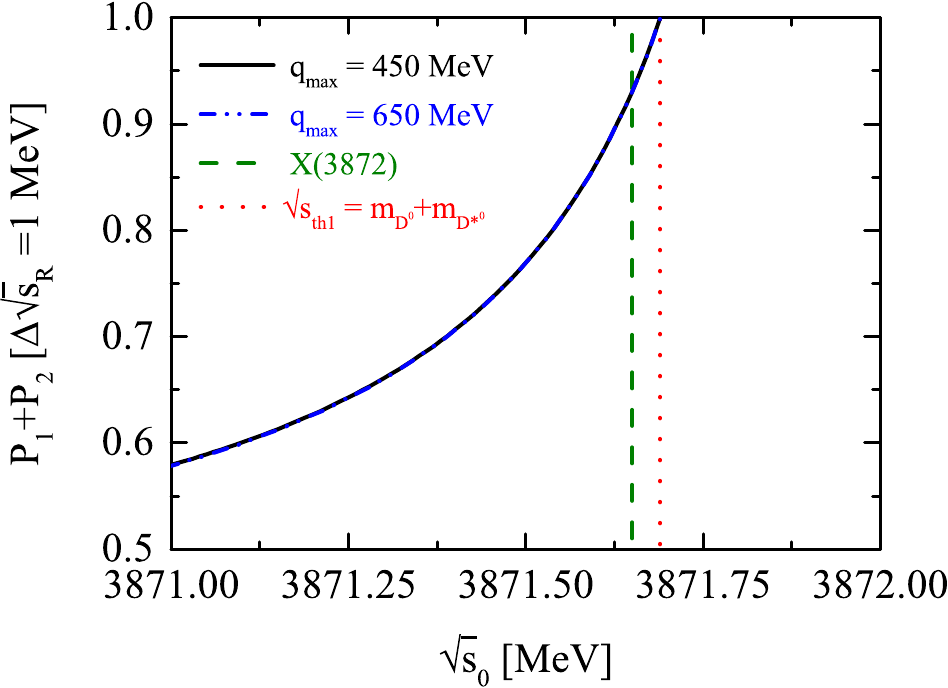}    
\caption{The labels are same as Fig.~\ref{FIG3}.}
    \label{FIG13}
\end{figure}

As mentioned before, we conduct now another test in which, in addition to the genuine state, we add the direct interaction between the mesons with a strength that does not bind by itself. The strength of this interaction is chosen from the local hidden gauge potential,  gauged by a factor such that the state would be bound with zero energy. As mentioned above, the strength of the potential and the regulator of the loop function $G$ of Eq.~(\ref{3_5}) are intimately related. We accomplish the previous task by multiplying Eq.~(\ref{3_17}) by $\beta=0.320$ for $q_\mathrm{max}=450$~MeV and  $\beta=0.485$ for $q_\mathrm{max}=650$~MeV. These values of $q_\mathrm{max}$ are in line with the value $420$~MeV demanded to get the experimental binding of the $T_{cc}(3875)$~\cite{Feijoo:2021ppq}. The results can be seen in Figs.~\ref{FIG11},~\ref{FIG13}. If we look now at Fig.~\ref{FIG11}, we see that for $\Delta\sqrt{s}_R=100$~MeV $P_1+P_2$ at the pole is around $1$, most of it coming from the $D^{*0}\Bar{D}^0$ channel. We should note that this number went up from $0.95$ in Fig.~\ref{FIG3} in the absence of any direct meson-meson interaction. In Fig.~\ref{FIG13}, we show the results for  $\Delta\sqrt{s}_R=1$~MeV. We see that $P_1+P_2$ at the pole is about $0.95$. But this number went up from $0.15-0.2$ in Fig.~\ref{FIG7} in the absence of direct meson-meson interaction. It is clear
that as soon as the extra meson-meson interaction is considered, the state becomes essentially molecular. 

As we can see, the presence of a reasonable direct meson-meson interaction has as a consequence a drastic increase in the molecular probability of the state.

Next we show in Table~\ref{value_a3_sR} the results that we obtain for $a$ and $r_0$ in this scenario. Comparing these results with those in Table~\ref{value_a1_sR}, we see that the consideration  of the direct meson interaction has also a drastic effect in the increase  of the scattering length and the decrease of the size of $r_0$. While for values of $\Delta\sqrt{s}_R$ of about $0.1$~MeV the value of $a_1$ and $r_{0,1}$ are still unacceptable. for $\Delta\sqrt{s}_R=1$~MeV, they can be acceptable with the current uncertainty in the experimental values. The discrepancy of $a$  with the value of $21.38$~fm with the  value of Eq.~(\ref{3_29}) is not significant in view of the present uncertainties in the binding of the $X(3872)$. However, we saw before that $P_1+P_2\sim0.95$ in this case. This means that at present, a mixed scenario with direct meson-meson interaction and a genuine state, with  molecular probabilities around $0.95$ can not be discarded. This scenario would be close to the one advocated in~\cite{Wang:2023ovj}. Yet, this is tied to the possible existence of a genuine state that prior to the dressing with the pion cloud has a mass extremely close to threshold, something that is not the case in ordinary tetraquark calculations. In order to see the differences with the scenario in which the state is purely molecular, generated exclusively from the meson-meson interaction we conduct a final test in the next subsection.

\subsection{Results from direct meson-meson interaction alone}

Now we demand that the $X(3872)$ is obtained from the meson-meson interaction without any contributions of the genuine state. 
This is accomplished by taking $\Tilde{g}^2=0$ and gauging to interaction of Eq.~(\ref{3_17}) with the factor $\beta$. 
This is accomplished by taking $\beta=0.324$ for $q_\mathrm{max}=650$~MeV and $\beta=0.494$ for $q_\mathrm{max}=450$~MeV 
(the value $\beta$ would be $\beta=0.537$ for $q_\mathrm{max}=420$~MeV used
in~{\cite{Feijoo:2021ppq}}). 
In this case the state is purely molecular~{\cite{Gamermann:2009uq}}, $P_1+P_2=1$, and we show in Table~\ref{value_a3} the results for $a$ and $r_0$. We can see that these values are very similar for any of the two values of  $q_\mathrm{max}$ used once we demand to obtain the bound state at the right energy. The small differences give  us a idea of the theoretical uncertainties that we can expect. The value for $a_1=22$~fm is in line with the one of Eq.~(\ref{3_29}) in view of the experimental uncertainties in the binding energy. It is also very similar to the value of $a_1$ obtained in Table~\ref{value_a3_sR} for $\Delta\sqrt{s}_R=1$~MeV when using the mixture of genuine state and direct meson-meson interaction. The radius $r_{0,1}$ is appreciably different $\sim0.5-(-)0.8$~fm versus $-2.30$~fm. It is thus clear
that an improvement in the measured value of $r_{0,1}$ can shed further light on the issue. On the other hand, there is extra information from $a_2$ and $r_{0,2}$. Curiously, these values in Table~\ref{value_a3} are very similar to those in Table~\ref{value_a3_sR} for $\Delta\sqrt{s}_R=1$~MeV, indicating that the crucial measurement are  those of $a_1$ and $r_{0,1}$, particularly the second. However, we should note that the values of Table~\ref{value_a3} for $a_2$ and $r_{0,2}$ are drastically different from those of Table~\ref{value_a1_sR} in the case of only the genuine state for small values of $\Delta\sqrt{s}_R$ where we would have chances of a small molecular component. All this is telling us that the precise values of $a_2,~r_{0,2},~a_2,~r_{0,2}$ are crucial to pin down the precise nature of the $X(3872)$. In this sense it is worth mentioning that this information is available for the $T_{cc}(3875)$~\cite{LHCb:2021auc,Mikhasenko:2022rrl,mishados:2023commu}  and this information was used in~\cite{Dai:2023cyo} to conclude that the $T_{cc}(3875)$ was purely molecular, with an uncertainty of the order of $2\%$. In this sense, there has been a recent revival concerning the relevance of $a,~r_0$ to determine the compositeness of states~\cite{Albaladejo:2022sux,Song:2022yvz}  improving on the Weinberg prescription~{\cite{Weinberg:1965zz}}, considering explicitly  
the range of the interaction. Definitely, the knowledge of $a, ~r_0$ for two coupled channels allows one to be more predictive as   proved in the case of $T_{cc}(3875)$ in~\cite{Dai:2023cyo}. We look forward to having these magnitudes measured with precision to give a definite answer to the problem.

 \section{Conclusions}
 We have conducted a thorough test on the molecular probability of the $X(3872)$ in the $D^0 \bar D^{*0}$ and  $D^+ D^{*-}$ components. We do three exercises. First we start from a genuine, nonmolecular state, which however has to couple to the former components to be observable in these channels, as is the case experimentally. Then we force the state to produce a bound state at a certain energy. We show that the state gets unavoidably dressed with the meson-meson components, to the point that the molecular probability becomes exactly unity when the binding approaches the $D^0 \bar D^{*0}$ threshold, with this channel acquiring most of the probability compared to the charged channel $D^+ D^{*-}$, eventually acquiring all of it in the limit of zero binding. Yet, there is the issue of how fast this probability goes to unity and we study this issue as a function of the bare mass of the genuine state prior to its dressing with the meson-meson components. We observe then that if the bare mass of the genuine state is very close to threshold, the raise of the molecular probability to unity occurs at even smaller distances to this threshold, to the point that for the experimental mass of the $X(3872)$ state, the molecular probability can be very small and we would have essentially a nonmolecular state. This is the first conclusion of the paper, that the binding energy by itself does not determine the nature of the  $X(3872)$ state. Yet, in this case we also observe that the scattering length of the $D^0 \bar D^{*0}$ channel becomes very small and the effective range grows up to values bigger than $500$~fm, in flagrant disagreement
with present experimental values.  The different cases studied allow us to conclude that if one starts with a pure genuine state, without consideration of any direct interaction of the meson-meson components and we force it to be responsible for the binding of the  $X(3872)$ state, the state becomes essentially pure molecular at the end. We definitely rule out this scenario.\\

    The next test consists in including a direct interaction between the meson-meson components. This interaction exists and is calculated by many groups independently. We take a reasonable interaction  provided by vector exchange, but scaled such that by itself does not produce binding of the meson-meson components. What we observe is that, as soon as a reasonable direct meson-meson interaction is considered, the molecular probability is drastically increased. We find however, that this kind of hybrid picture is not forbidden by present data of the scattering length and effective range, but even then the molecular probability is of the order of $95 \%$.\\

Finally, we also conduct a test using a pure molecular picture in which there is no genuine state.  In this case, by construction, the molecular probability is unity but we determine the values of the scattering length and effective range and see that, while certainly compatible with present experimental values, they differ appreciably from the hybrid scenario discussed above. We also point out the relevance of the scattering length and effective range for the $D^+ D^{*-}$ channel, that has not been given attention so far, neither theoretically nor experimentally, and conclude that a determination of these magnitudes together with more precise values of $a$ and $r_0$ for the $D^0 \bar D^{*0}$   channel will be extremely useful in the future to further pin down the molecular probability of the $X(3872)$. While with present data we can certainly rule out a picture in which the nonmolecular component of the $X(3872)$ is dominant, the precise value of the molecular components will have to wait for more precise measurements of the scattering length and effective range for the $D^0 \bar D^{*0}$ and $D^+ D^{*-}$ channels.

\begin{table}[H]
\footnotesize
\centering
 \caption{The value of scattering length $a$ and $r_0$ at threshold in  different $q_\mathrm{max}$.}\label{value_a1_sR}
\setlength{\tabcolsep}{6pt}
\begin{tabular}{ccccc|cccc}
\hline
\hline
\multirow{2}{*}{$\Delta \sqrt{s}_R[\mathrm{MeV}]$} &  \multicolumn{4}{c|} {$q_\mathrm{max}=450~$MeV } & \multicolumn{4}{c}{ $q_\mathrm{max}=650~$MeV}   \\
  \cline{2-9}& $a_1[\mathrm{fm}]$ &  $r_{0,1}[\mathrm{fm}]$ & $a_2[\mathrm{fm}]$ &  $r_{0,2}[\mathrm{fm}]$ & $a_1[\mathrm{fm}]$ &  $r_{0,1}[\mathrm{fm}]$  & $a_2[\mathrm{fm}]$ &  $r_{0,2}[\mathrm{fm}]$\\
\hline
$0.1$ &$1.42$&$-663.61$&$0.0073 - i~0.00003 $&$-664.79 - i~1.56$ &$0.954$ &$-1011.3$&$0.0048 - i~0.00002 $&$-1014.0 - i~1.56$          \\
$0.3$ &$3.16$ &$-273.51$&$0.0176 - i~0.00020 $&$-273.04 - i~1.56$  &$2.181$ &$-416.86$&$0.0116 - i~0.00009 $  &$-417.03 - i~1.56$       \\
$1$  & $7.48$ &$-89.71$ & $0.0530 - i~0.00180 $ & $-88.46 - i~1.56$  & $5.544$  & $-136.78$   & $0.0350 - i~0.00078 $ & $-135.77 - i~1.56$        \\
$2$ &$11.09 $&$-45.95$&$0.1014 - i~0.00660 $&$-44.52 - i~1.56$ &$8.760$&$-70.098$&$0.0674 - i~0.00292  $&$-68.81 - i~1.56$         \\
$5$ &$15.80 $&$-18.86$&$0.2305 - i~0.03475 $&$-17.31 - i~1.56 $&$13.67$&$-28.816$&$0.1571 - i~0.01597   $&$-27.35 - i~1.56 $       \\
$10$ &$18.45$ &$-9.68$&$0.3957 - i~0.10756  $&$-8.10 - i~1.56 $ &$16.87 $&$-14.837$&$0.2827 -i~0.05290 $&$-13.31 - i~1.56  $       \\
$20$ &$20.16$ &$-5.07$&$0.5902 - i~0.26910   $&$-3.47 - i~1.56 $&$19.11 $&$-7.8049$&$0.4593 - i~0.14915   $&$-6.25 - i~1.56  $    \\
$50$  &$21.35$&$-2.29$&$0.7558 - i~0.58190   $&$-0.68 - i~1.56 $ &$20.79$ &$-3.5725$&$0.6801 - i~0.39616   $ &$-2.00 - i~1.56$   \\
$70$  &$21.59$&$-1.76$&$ 0.7761 - i~0.68790  $ &$-0.15 - i~1.56$&$21.14$ &$-2.7652$&$0.7296 - i~0.50085 $&$-1.19 - i~1.56$       \\
$100$ &$ 21.78$  &$-1.37 $& $0.7818 - i~0.78157 $&$0.25 - i~1.56  $&$ 21.41 $ &$-2.1595 $& $0.7611 - i~0.60330 $&$-0.58 - i~1.56  $        \\
   \hline
   \hline
   \end{tabular}
\end{table}

\begin{table}[H]
\footnotesize
\centering
 \caption{The value of scattering length $a$ and $r_0$ at threshold in  different $q_\mathrm{max}$.}\label{value_a3_sR}
\setlength{\tabcolsep}{8pt}
\begin{tabular}{ccccc|cccc}
\hline
\hline
\multirow{2}{*}{$\Delta \sqrt{s}_R[\mathrm{MeV}]$} &  \multicolumn{4}{c|} {$q_\mathrm{max}=450~$MeV } & \multicolumn{4}{c}{ $q_\mathrm{max}=650~$MeV}   \\
  \cline{2-9}& $a_1[\mathrm{fm}]$ &  $r_{0,1}[\mathrm{fm}]$ & $a_2[\mathrm{fm}]$ &  $r_{0,2}[\mathrm{fm}]$ & $a_1[\mathrm{fm}]$ &  $r_{0,1}[\mathrm{fm}]$  & $a_2[\mathrm{fm}]$ &  $r_{0,2}[\mathrm{fm}]$\\
\hline
$0.1$ &$15.60$&$-24.97$&$0.7068 - i~1.116 $&$1.17 - i~1.56   $&$ 15.84  $&$-25.74  $&$0.7476 - i~1.011    $&$   0.82 - i~1.56 $       \\
$0.3$ & $19.65  $&$ -7.13 $&$0.7060 - i~1.118   $&$ 1.16 - i~1.56   $&$  19.50$&$ -7.55 $&$ 0.7470 - i~1.012  $&$  0.81 - i~1.56 $      \\
$1$  & $21.38  $&$-2.30  $&$ 0.7024 - i~1.125  $&$ 1.14 - i~1.56   $&$  21.23 $&$ -2.64 $&$0.7448 - i~1.019   $&$ 0.78 - i~1.56 $    \\
$2$   &$21.79 $&$ -1.35 $&$ 0.6957 - i~1.139   $&$ 1.08 - i~1.56  $&$ 21.64  $&$ -1.68 $&$ 0.7406 - i~1.032  $&$  0.72 - i~1.56 $       \\
$5$   &$ 22.05 $&$ -0.81 $&$ 0.6394 - i~1.232  $&$ 0.23 - i~1.56   $&$  21.90$&$-1.13  $&$0.7035 - i~1.128   $&$ -0.11 - i~1.56 $      \\
$10$  &$ 22.13  $&$ -0.63 $&$ 0.7818 - i~0.780  $&$ -3.62 - i~1.56   $&$ 21.98 $&$ -0.94 $&$0.7767 - i~0.693   $&$ -4.32 - i~1.56 $        \\
$20$  &$ 22.17  $&$-0.54  $&$ 0.7514 - i~0.998  $&$ 0.92 - i~1.56   $&$ 22.02  $&$ -0.85 $&$ 0.7731 - i~0.898  $&$  0.56 - i~1.56 $       \\
$50$  &$ 22.20 $&$ -0.48 $&$ 0.7410 - i~1.031  $&$ 1.12 - i~1.56   $&$  22.05 $&$ -0.80 $&$0.7677 - i~0.930   $&$  0.77 - i~1.56 $      \\
$70$  &$ 22.21 $&$ -0.47 $&$ 0.7396 - i~1.035  $&$  1.14 - i~1.56  $&$  22.05 $&$-0.79  $&$0.7669 - i~0.934   $&$  0.79 - i~1.56 $       \\
$100$  &$ 22.21 $&$-0.47  $&$ 0.7385 - i~1.038  $&$ 1.15 - i~1.56 $&$  22.06   $&$ -0.78 $&$  0.7663 - i~0.937   $&$0.80 - i~1.56  $  \\
   \hline
   \hline
   \end{tabular}
\end{table}

\begin{table}[H]
\footnotesize
\centering
 \caption{The value of scattering length $a$ and $r_0$ at threshold in  different $q_\mathrm{max}$.}\label{value_a3}
\setlength{\tabcolsep}{30pt}
\begin{tabular}{ccccc}
\hline
\hline
%\multirow{2}{*}{$q_\mathrm{max}[\mathrm{MeV}]$} &  \multicolumn{4}{c|} {$s_R=(m_D+m_{D^*}+100)^2$} & \multicolumn{4}{c}{ $s_R=(m_D+m_{D^*}+1)^2$}   \\
  $q_\mathrm{max}[\mathrm{MeV}]$& $a_1[\mathrm{fm}]$ &  $r_{0,1}[\mathrm{fm}]$ & $a_2[\mathrm{fm}]$ &  $r_{0,2}[\mathrm{fm}]$\\
\hline
$450$ &$22.22 $ & $-0.449$ & $0.736  - i~1.04  $ & $1.17 - i~1.56$ \\
$650$ & $22.07 $ & $-0.763 $& $0.765 - i~0.94 $& $0.82 - i~1.56$\\
   \hline
   \hline
   \end{tabular}
\end{table}

\newpage

\section{acknowledgements}

This work is partly supported by the National Natural Science Foundation of China under Grants Nos. 12175066,
11975009, 12247108 and the China Postdoctoral Science Foundation under Grant No. 2022M720359. This work is
also partly supported by the Spanish Ministerio de Economia y Competitividad (MINECO) and European FEDER
funds under Contracts No. FIS2017-84038-C2-1-P B, PID2020-112777GB-I00, and by Generalitat Valenciana under
contract PROMETEO/2020/023. This project has received funding from the European Union Horizon 2020 research
and innovation programme under the program H2020-INFRAIA-2018-1, grant agreement No. 824093 of the STRONG-2020 project. This research is also supported by the Munich Institute for Astro-, Particle and BioPhysics (MIAPbP)
which is funded by the Deutsche Forschungsgemeinschaft (DFG, German Research Foundation) under Germany’s
Excellence Strategy-EXC-2094 -390783311.

\bibliography{refs.bib}
\end{document}